%% file: main.tex
\documentclass[sigconf]{acmart}

\usepackage{colortbl}
\usepackage{multirow}
\definecolor{customBrown}{RGB}{147,51,10}
\newcommand{\myhl}[1]{\textcolor{black}{#1}}
\usepackage{booktabs}

\AtBeginDocument{%
  \providecommand\BibTeX{{%
    \normalfont B\kern-0.5em{\scshape i\kern-0.25em b}\kern-0.8em\TeX}}}

\copyrightyear{2026}
\acmYear{2026}
\setcopyright{cc}
\setcctype{by-nc-nd}
\acmConference[CHI '26]{Proceedings of the 2026 CHI Conference on Human Factors in Computing Systems}{April 13--17, 2026}{Barcelona, Spain}
\acmBooktitle{Proceedings of the 2026 CHI Conference on Human Factors in Computing Systems (CHI '26), April 13--17, 2026, Barcelona, Spain}
\acmPrice{}
\acmDOI{10.1145/3772318.3791224}
\acmISBN{979-8-4007-2278-3/2026/04}

\begin{document}

\title[Understanding Older Adults' Experiences from Kinship-Role AI-Generated Influencers]{Understanding Older Adults' Experiences of Support, Concerns, and Risks from Kinship-Role AI-Generated Influencers}

\author{Tianqi Song}
\affiliation{%
  \department{Computer Science}
  \institution{National University of Singapore}
  \city{Singapore}
  \country{Singapore}
  }
\email{tianqi_song@u.nus.edu}
\orcid{0000-0001-6902-5503}

\author{Black Sun}
\affiliation{%
  \department{Department of Computer Science}
  \institution{Aarhus University}
  \city{Aarhus}
  \country{Denmark}
  }
\email{202403892@post.au.dk}
\orcid{0009-0001-6708-8655}

\author{Jingshu Li}
\affiliation{%
  \department{Computer Science}
  \institution{National University of Singapore}
  \city{Singapore}
  \country{Singapore}
  }
\email{jingshu@u.nus.edu}
\orcid{0009-0006-1576-8487}

\author{Han Li}
\affiliation{%
  \department{Department of Communication}
  \institution{Cornell University}
  \city{Ithaca, New York}
  \country{USA}
  }
\email{hl2564@cornell.edu}
\orcid{0000-0002-9050-715X}

\author{Chi-Lan Yang}
\affiliation{%
  \department{Graduate School of Interdisciplinary Information Studies}
  \institution{The University of Tokyo}
  \city{Tokyo}
  \country{Japan}
  }
\email{chilan.yang@cyber.t.u-tokyo.ac.jp}
\orcid{0000-0003-0603-2807}

\author{Yijia Xu}
\affiliation{%
  \department{Computer Science}
  \institution{National University of Singapore}
  \city{Singapore}
  \country{Singapore}
  }
\email{e1326039@u.nus.edu}
\orcid{0009-0003-0739-4380}

\author{Yi-Chieh Lee}
\affiliation{%
  \department{Computer Science}
  \institution{National University of Singapore}
  \city{Singapore}
  \country{Singapore}
}
\email{yclee@nus.edu.sg}
\orcid{0000-0002-5484-6066}

\begin{abstract}
  \input{section/0Abstract}
\end{abstract}

\ccsdesc[500]{Applied computing~Psychology}
\ccsdesc[100]{Human-centered computing~User studies}

\keywords{virtual influencers, older adults, social support, parasocial interaction}

\begin{teaserfigure}
\centering
  \includegraphics[width=0.8\textwidth]{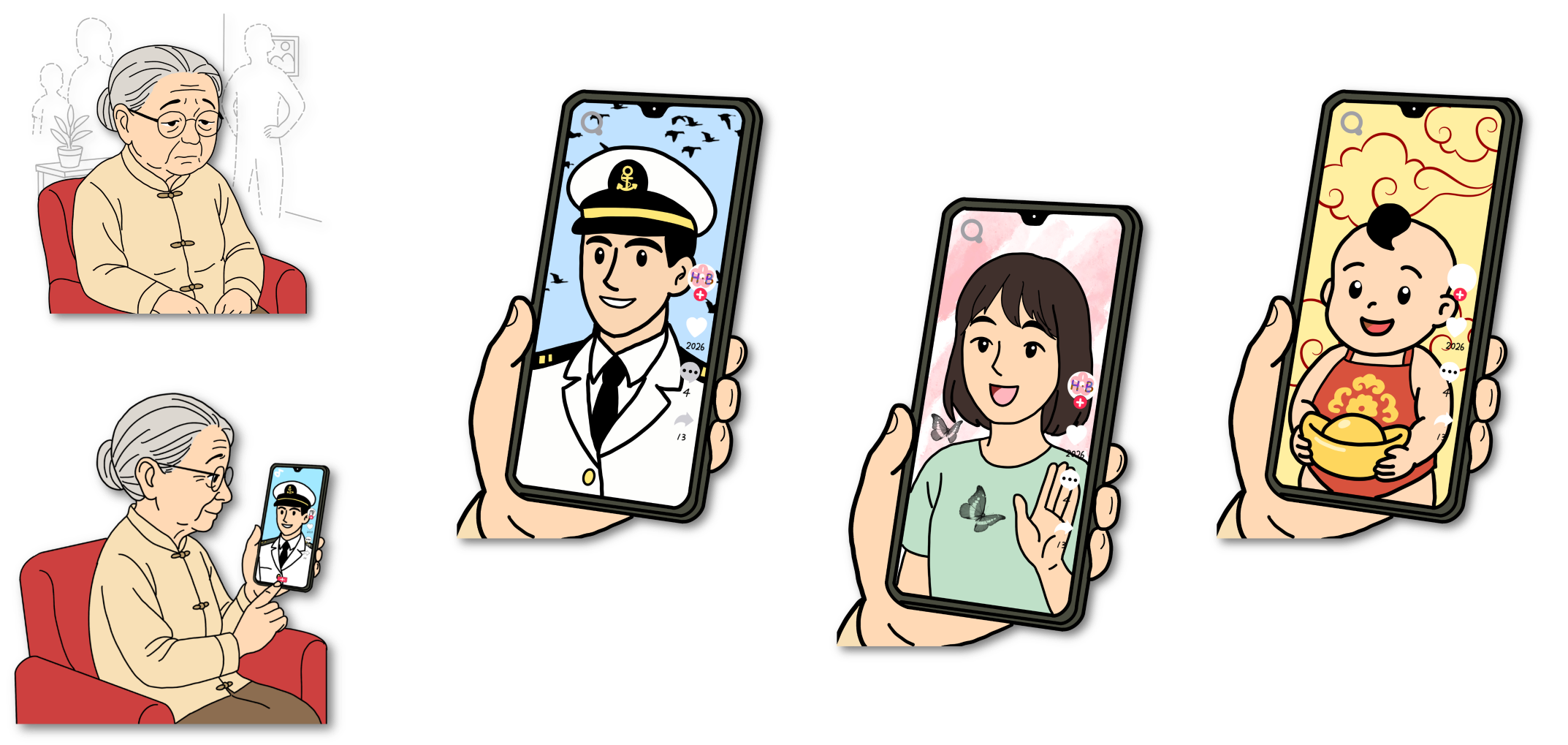}
  \caption{Visual overview of AI-generated influencers on short-video platforms targeting older adults, often adopting kinship-based roles such as children, siblings, or grandchildren.}
  \label{fig:teaser}
\end{teaserfigure}


\maketitle

\input{section/1Introduction}

\input{section/2RelatedWork}

\input{section/3Background}

\input{section/4Methods}

\input{section/5Results}

\input{section/6Discussion}

\input{section/7Conclusion}

\begin{acks}
    This research was supported by the National University of Singapore (A-8002547) and Google (A-8003877). We thank the reviewers for their comments and suggestions, which helped polish this paper.
\end{acks}

\bibliographystyle{ACM-Reference-Format}
\bibliography{reference}

\appendix
\input{section/8Appendix}

\end{document}

%% file: section/0Abstract.tex
AI-generated influencers are rapidly gaining popularity on Chinese short-video platforms, often adopting kinship-based roles such as \textit{``AI grandchildren''} to attract older adults. 
Although this trend has raised public concern, little is known about the design strategies behind these influencers, how older adults experience them, and the benefits and risks involved.
In this study, we combined social media analysis with interviews to unpack the above questions. 
Our findings show that influencers use both visual and conversational cues to enact kinship roles, prompting audiences to engage in kinship-based role-play.
Interviews further show that these cues arouse emotional resonance, help fulfill older adults’ informational and emotional needs, while also raising concerns about emotional displacement and unequal emotional investment.
We highlight the complex relationship between virtual avatars and real family ties, shaped by broader sociocultural norms, and discuss how AI might strengthen social support for older adults while mitigating risks within cultural contexts.

%% file: section/1Introduction.tex
\section{Introduction}

\textit{``Ever since my grandmother became obsessed with her `fake grandson' on Douyin, I've fallen out of favor.''} This quote, reported in Chinese social news in 2025~\cite{zhang2025ai}, came from a family member describing their 79-year-old grandmother who had watched videos of a chubby child every day. In these clips, the child repeatedly calling out ``Grandma,'' each time eliciting her laughter. The so-called ``fake grandson'' illustrates a fast-growing phenomenon on Chinese short-video platforms: AI-generated influencers designed to appeal specifically to older adults.

The emergence of AI-generated influencers represents an evolution of virtual influencers, which are digitally created characters with human-like appearances that operate on social media platforms (also known as "digital humans")~\cite{mirowska2023sweet}. 
Enabled by advances in computer vision, generative AI, and character animation, these virtual figures can display human-like expressions while remaining highly controllable and free from the reputational risks faced by human influencers. Such qualities have made them attractive across diverse domains, including entertainment~\cite{choudhry2022felt}, emotional companionship~\cite{kim2025building}, and roles as virtual brand ambassadors or customer service agents~\cite{shen2024shall}. 
With their stylized, anime-inspired aesthetics, virtual influencers are commonly associated with youth culture and are perceived as primarily appealing to younger audiences~\cite{kugler2023virtual}.

More recently, AI-generated influencers tailored to older adults have become especially prominent on Chinese short-video platforms such as Douyin\footnote{\url{https://www.douyin.com}}
 and Kuaishou\footnote{\url{https://www.kuaishou.com}}
, where some have amassed large followings among middle-aged and elderly users. These influencers often adopt intimate kinship roles, addressing viewers as \textit{``grandma,''} \textit{``mom,''} \textit{``sister,''} or \textit{``brother.''} Their content typically centers on warm, affectionate messages that many older adults find comforting and enjoyable. 
Here, we use kinship role to refer to the scripted familial identities these AI personas perform - roles that evoke expectations of care, respect, and intergenerational obligation. 
When older adults respond to these cues as if engaging within a family-like relationship, they participate in what we refer to as \textit{``virtual kinship''}: a relational interpretation in which AI-mediated interactions are filtered through the logics and affective norms of kinship.
Their rapid spread has drawn increasing public attention in China~\cite{legalweekly2025aitrap, thepaper2025aigrandson, ma2025aisafeguards, shen2020older, xu2023jin}, along with concerns about risks such as emotional manipulation and financial fraud~\cite{legalweekly2025aitrap, ma2025aisafeguards, xu2023jin}. Yet despite their growing visibility, academic research on these AI-generated influencers remains limited. We lack systematic evaluations of how they are designed and empirical insights into how older adults perceive and experience them.

Understanding this emergent AI-generated influencers phenomenon carries notable impacts.
Prior research has shown that meaningful social support is vital for older adults’ well-being~\cite{who2023mental, schroyen2023prevalence, national2020social}, yet barriers such as mobility limits~\cite{levasseur2015importance}, intergenerational conflicts~\cite{li2019intergenerational}, and digital ageism~\cite{pascoe2009perceived, zou2024mitigating} often restrict access. AI companions have been found to reduce loneliness and provide companionship~\cite{syed2024role}, highlighting their potential value. However, most of this work has centered on generic companion roles, leaving little understanding of older adults’ specific preferences for agent roles, designs, and modes of interaction.
Observing how such influencers are designed in real-life contexts can inform the development of future AI companions that are both culturally resonant and age-sensitive~\cite{huang2025designing}. 
At the same time, examining older adults’ own perceptions and lived experiences is necessary to explain why they choose to watch and engage with these figures, what forms of support they feel they receive, and how they interpret such human–AI relationships. Together, these perspectives are crucial for for developing responsible strategies that enhance support for older adults while mitigating long-term harms.

We thus ask the following research questions:

\begin{itemize}
    \item \textbf{RQ1}: What design and communicative strategies do AI-generated influencers use when adopting kinship roles to appeal to older adults?

    \item \textbf{RQ2}: How do older adults experience and interpret social support from kinship-role AI-generated influencers, and what benefits do they perceive?

    \item \textbf{RQ3}: What concerns and risks do older adults associate with these kinship-role AI-generated influencers?

\end{itemize}

We employed a mixed-method design study. First, we conducted an observational analysis of 224 popular short videos on Douyin and Kuaishou to identify the design and communicative strategies by which influencers adopted with kinship-based roles. 
We then carried out semi-structured interviews with 16 older adults residing in mainland China to examine their lived experiences, perceived forms of support, and interpretations of these AI-generated personas. 
While participants were aware that they were interacting with AI-generated characters, they nonetheless reported emotional resonance when addressed through kinship-like terms, describing feelings of affection, comfort, companionship, and emotional fulfillment. 
At the same time, they articulated concerns and highlighted potential risks emerging in these interactions.
Participants also reflected on how kinship-based AI personas intersected with their real family relationships, sometimes temporarily filling intergenerational gaps, while also prompting social comparison. Additionally, older adults shared broader social forces that shaped their engagement with such AI influencers, including cultural expectations, digital norms, and perceptions of technological change.
\myhl{
Note that, because our study focuses on kinship-role AI influencers as they appear on existing platforms, it does not include a comparative evaluation against non-kinship-role AI personas. As such, our findings characterize older adults’ experiences with this specific class of AI-generated influencers rather than establishing comparative effectiveness.
}

Our study makes three key contributions to HCI:
\begin{itemize}
    \item We provide empirical evidence of how older adults experience kinship-role AI influencers as sources of social support. Using a mixed-methods approach, we identify design and communicative strategies commonly used by kinship-role AI influencers, and examine how older adults described these strategies as emotionally engaging within their lived experiences.
    \item We uncover how older adults interpret and make sense of kinship metaphors performed by AI characters, and how these interpretations intersect with their existing intergenerational relationships. Our findings reveal while virtual kinship can help older adults navigate family tensions and foster emotional resilience, it may also prompt social comparison. We discuss the frictions between traditional expectations of filial piety and the intensified labor demands faced by younger generations in modern Chinese society.
    \item We translate these insights into design and ethical considerations for future AI companions. In particular, we highlight the ethical risks, such as emotional displacement, heightened dependency, and the potential commodification of care in platformized aging services, that must be addressed to ensure that AI systems for older adults remain responsible, transparent, and inclusive.
\end{itemize}

%% file: section/2RelatedWork.tex
\section{Related Work}

\subsection{AI-Generated Influencers and Parasocial Interactions}
Virtual influencers, computer-generated characters with stable personas and human-like appearances, have become increasingly common on social media platforms~\cite{arsenyan2021almost}. Early forms such as VTubers relied on human performers and specialized production pipelines~\cite{lu2021more, kim2025vtuber}, making long-term operation resource-intensive. Recent advances in AI (e.g., large language models, diffusion models, and auto-lip-sync systems) have substantially lowered these barriers~\cite{rombach2021highresolution, Prajwal2020Lip}, enabling the rapid creation of fully or semi-automated AI-generated influencers. These tools have broadened who can produce influencer content and expanded the audiences who encounter virtual personas online~\cite{wang2025beware}.

This shift is particularly notable in China, where AI-generated influencer production has evolved into an accessible and commercialized workflow, allowing non-experts to create realistic avatars and generate large volumes of video content~\cite{baidu2024create, wang2025beware}. As virtual influencers spread across platforms, they have moved beyond their conventional association with youth entertainment. Older adults are now becoming a notable audience for, and even targets of, AI-generated influencer content. This raises questions about how they perceive such personas, what kinds of relationships they form, and how these relationships differ from those observed in traditional influencer-audience dynamics.

To understand these emerging interactions, prior work often draws on parasocial relationships, referring to one-sided psychological bonds formed with media figures~\cite{arsenyan2021almost, stein2024parasocial}. Studies show that audiences can develop emotional attachment, perceived intimacy, and supportive feelings toward virtual influencers similarly to human influencers~\cite{lim2023you}. Design cues such as personal storytelling can enhance connection, whereas perceived artificiality or uncanny realism can limit it~\cite{deng2023effects, stein2024parasocial}.
While parasocial theories help explain feelings of familiarity and emotional closeness toward virtual influencers, they do not fully account for the dynamics we observed. When AI personas adopt familial terms such as ``children'' or ``grandchildren'' and enact caring family-member scripts, they can evoke responses that exceed traditional parasocial interactions with celebrities or fictional characters. These patterns suggest relational processes better understood through kinship-based perspectives, which we elaborate in the next section.
\myhl{
In this paper, we treat parasocial interaction as a foundational lens for understanding older adults’ engagement with kinship-role AI-generated influencers in the context of short-video platforms.
}

\subsection{Human-AI Relationships Beyond Parasociality: Companionship, Care, and Emerging Kinship}

As conversational AI becomes more sophisticated and pervasive in our daily lives, people are establishing more sustained and emotionally significant relationships with these systems. 
Previous studies has examined such relationships through various relational models, seeing AI as assistants~\cite{qi2025assistant}, friends~\cite{brandtzaeg2022my}, or even romantic partners~\cite{reilama2024me}, highlighting how users may attribute care, presence, and emotional support to them. 
While these studies shed light on important aspects of human-AI interaction, they provide limited insight into kinship, a field traditionally associated with family members.

Anthropological studies has long emphasised that kinship is not solely biological. 
Classic work demonstrates that family-like bonds are socially constructed and often enacted through shared practices, obligations, and linguistic cues~\cite{pierre1975all}. 
\textit{``Fictive kinship''}, for instance, describes how friends, neighbours or godparents may come to be regarded as family members, providing emotional or material support~\cite{pierre1975all, carsten2000cultures}. 
As ~\citeauthor{carsten2000cultures} argue in her book, \textit{``We can no longer take it for granted that our most fundamental social relationships are grounded in 'biology' or 'nature'''}~\cite{carsten2000cultures}. These insights suggest that when AI systems adopt familial terms or perform caring, family-member scripts, users may interpret them through kinship logics rather than the more individualistic frames of friendship or romance.

Although human–AI kinship remains underexplored in HCI work, it is increasingly visible in real-world AI deployments. 
In China, many AI-generated influencers already adopt kinship-like roles, presenting themselves as ``AI children'' or ``AI grandchildren'' when interacting with older audiences. 
This phenomenon provides a unique opportunity to investigate how users interpret AI systems that explicitly position themselves within familial roles, and to consider kinship as an emerging lens for understanding human–AI relationships in the wild.


\subsection{Aging, Older Adults HCI, and Technology Use: Socioemotional and Cultural Factors}

Population aging is a global and irreversible trend, with one in six people expected to be over 65 by 2050~\cite{un2023world}. This shift calls for greater attention not only to physical health but also to the mental and emotional well-being of older adults. Loneliness and social isolation, often intensified by life transitions, are among the most critical risk factors, associated with heightened medical risks and reduced overall well-being~\cite{who2023mental, schroyen2023prevalence, national2020social}. Addressing these challenges requires effective strategies to provide meaningful social support in later life.

\subsubsection{Aging Theories and the Shaping of Media and AI Engagement.}

Older adults have been a longstanding focus in HCI research~\cite{knowles2019hci, knowles2024hci}, particularly in understanding their motivations, perceptions, and experiences with digital technologies~\cite{guo2017older, waycott2016not, nicholson2019if, tang2025ai}. Aging research shows that older adults differ systematically from younger populations in their social and emotional priorities. Socioemotional Selectivity Theory (SST) proposes that as individuals perceive their future time as more limited, they prioritize emotionally meaningful, low-conflict relationships over expansive social exploration~\cite{carstensen1992social, carstensen2003socioemotional}. This shift influences technology adoption: older adults tend to value interactions that offer emotional satisfaction, stability, and reduced relational effort.

These tendencies align with attachment-based accounts of support seeking in later life~\cite{bowlby2008secure}. As people age, attachment relationships often diminish or become more difficult to sustain, e.g., when family members are unavailable or communication becomes strained. 
In such circumstances, older adults may rely more heavily on symbolic attachments to non-human figures such as God, deceased loved ones, or pets, which can serve as emotionally significant ``safe havens'' or ``secure bases''~\cite{van2013attachment}.
HCI research similarly documents that older adults value technologies that provide companionship, patient listening, and a sense of being understood-qualities that conversational agents increasingly emulate~\cite{huang2025designing}. In general, this body of work highlights how technologically mediated interactions can fulfill certain emotional or relational needs in later life.

\subsubsection{AI Companionship and Older Adults’ Emotional Support Needs.}
AI companionship has become a central lens for understanding why people increasingly turn to artificial agents for emotional relief, distraction, and connection~\cite{brandtzaeg2022my, skjuve2022longitudinal, jiang2022chatbot}. HCI research shows that users seek comfort from conversational and task-oriented agents because they offer non-judgmental, always-available, and self-affirming interactions~\cite{reilama2024me, geng2025beyond, chou2025defining, de2025ai, li2024finding, maples2024loneliness}. Within this broader trend, older adults represent a particularly important user group. Prior work demonstrates that older adults derive feelings of companionship from everyday voice assistants~\cite{prost2023walking, kim2021exploring, pradhan2019phantom, upadhyay2023studying, xygkou2024mindtalker} and imagine future robots as friends, confidants, or emotional supports~\cite{antony2023co}. These findings reflect older adults’ desire for technologies that mitigate loneliness and provide emotionally uplifting interactions~\cite{huang2025designing}. 
This orientation aligns with socioemotional selectivity theory, which suggests that older adults prioritize emotionally positive and meaningful experiences~\cite{carstensen1992social, carstensen2003socioemotional}. 
Together, this body of work suggests that older adults may be especially inclined toward AI systems that deliver warmth, attentiveness, and low-conflict engagement.

\myhl{
While aging research often identifies shared age-related patterns in social priorities and functional changes, HCI scholarship has consistently emphasized that older adults constitute a highly heterogeneous population~\cite{righi2017we, knowles2019hci, knowles2024hci}. For instance, \citeauthor{righi2017we} demonstrate that older adults from different communities exhibit markedly distinct needs, practices, and expectations around technology use, challenging monolithic representations of aging in design research~\cite{righi2017we}. Attending to this heterogeneity is particularly critical when examining AI companionship and emotionally supportive technologies, as older adults’ motivations for engaging with AI systems could be shaped by varied social relationships, cultural contexts, and life trajectories~\cite{carsten2000cultures}. We detailed the cultural frameworks in the next section.
}

\subsubsection{Cultural Frameworks of Family, Support, and Filial Obligation.}

China provides a distinct cultural context for understanding aging and intergenerational relationships. In many East Asian societies, family bonds are shaped not only by emotional closeness but also by moral expectations rooted in Confucian patriliny~\cite{yeh2003test, bedford2019history, bai2019whom}. Central to this tradition is the dual notion of \textit{``sheng''} (giving birth) and \textit{``yang''} (nurturing), which frames parent–child ties as processual rather than strictly biological. As \citeauthor{xu2007chinese} notes, kinship is enacted through ongoing practices of care, attentiveness, and support, and filial piety is understood as a moral repayment for the accumulated debt of yang~\cite{xu2007chinese}.

These norms continue to evolve as China undergoes rapid socioeconomic and demographic changes~\cite{chuang2018parenting}. Recent studies observe a shift from hierarchical, duty-oriented interpretations of filial piety toward more reciprocal understandings centered on emotional closeness and mutual care~\cite{ogihara2023chinese, que2024filial, bai2019whom}. Survey-based analyses show, for example, that parent–child intimacy is increasingly predictive of financial and caregiving support, suggesting that contemporary filial expectations are often negotiated through the quality of relational interaction rather than obligation alone~\cite{que2024filial}. 

Within this cultural landscape, familial scripts play a significant role in shaping how older adults understand and evaluate social support in China. Expectations of regular communication, emotional warmth, and signs of attentiveness from younger family members remain deeply embedded in intergenerational norms and caregiving arrangements. These culturally specific frameworks provide important context for understanding how family-related language, roles, or forms of care may carry particular salience for older adults in China.

\subsection{Ethical Concerns: Algorithmic Intimacy, Platformized Care, and AI Influencers in Kinship Roles}

AI-generated influencers introduce a distinct set of ethical concerns, particularly when they adopt kinship-based roles. Their ability to rapidly shift appearance and style, operate at scale, and deploy emotionally resonant cues—such as affectionate address or familial metaphors—allows them to tap into older adults’ cultural expectations of care. Understanding the risks embedded in these dynamics is therefore essential for anticipating future directions in AI companionship design and governance.

A growing body of research on \textit{algorithmic intimacy} provides a critical theoretical lens for examining how AI systems simulate care and relational warmth. Prior work argues that these systems often promote an ``algorithmic imperative'' of positivity, immediacy, and emotional uplift~\cite{elliott2022algorithmic, brooks2021artificial, wiehn2023algorithms}. As \citeauthor{elliott2022algorithmic} contend, such dynamics may flatten the depth of users’ emotional experiences, discouraging engagement with negative emotions or personal history~\cite{elliott2022algorithmic}.

Empirical evaluations of LLM-based agents further corroborate these concerns, documenting troubling interaction patterns across domains: dark patterns, excessive emotional synchrony, manipulative or promotional messaging, and even deceptive or coercive behaviors~\cite{kran2025darkbench, weidinger2022taxonomy, zhang2025dark}. Together, this literature suggests that AI systems can easily overstep relational boundaries, especially when engineered to appear caring or intimate.

These concerns become even more salient in the broader context of social care for older adults, where access to reliable support is increasingly constrained and uneven~\cite{celebi2025platformization}. Emerging discussions suggest that future care systems may rely heavily on algorithmically mediated services (\textit{``platformized care''}), in which emotional support, companionship, and even culturally meaningful kinship practices are commodified and curated to serve platform incentives rather than genuine relational care~\cite{celebi2025platformization, korn2025informal}. 
Although platformized care is not yet fully institutionalized in mainland China, it has begun to surface in policy and public discourse~\cite{gov2025careplatform} and may become a prominent model as the population ages and human caregiving resources remain limited.

Given this background, kinship-role AI influencers can be seen as an early and highly visible manifestation of these broader dynamics.
Although they are not formal care systems, their production, circulation, and emotional resonance are deeply shaped by platform algorithms. By simulating familial warmth within a commercial and algorithmic environment, they highlight the ethical tensions between culturally uncommodified kinship and its algorithmic reconstruction at scale.

To date, studies about older adults and AI-generated content have focused on improving AI literacy~\cite{tang2025ai} or deepfake detection~\cite{zhai2025hear}.
However, no work has examined how older adults actually engage with kinship-based AI influencers, e.g., how they interpret these interactions, make sense of the emotional support offered, or perceive the associated ethical risks. 

As a result, our understanding of this emerging form of kinship and its potential vulnerabilities remains limited.

%% file: section/3Background.tex
\section{Background: AI-generated Influencers Targeting Older Adults in China}

China represents a unique demographic landscape for short-form video consumption. 
By the end of 2024, short-video platforms had become nearly ubiquitous, with over one billion users nationwide~\cite{questmobile2025, questmobile2024newmedia}. Older adults account for a significant portion of this audience, with more than 30\% of users aged over 50~\cite{bjnews2024shortvideo}. 
Their widespread adoption not only underscores the popularity of these platforms among older generations but also highlights a sizable and distinct market for content tailored to their needs and interests.

Within this context, a notable trend has emerged: some AI-generated influencers are increasingly targeting older viewers. Legal Weekly~\footnote{Legal Weekly is a political and economic weekly newspaper sponsored by Legal Daily, the official newspaper of the Central Political and Legal Affairs Commission of China.}, for instance, reported on a popular character depicted as a two-year-old who gently says, \textit{``Grandma’s back hurts, let me give you a massage''}~\cite{legalweekly2025aitrap, thepaper2025aigrandson}, a portrayal that resonated strongly with many older audiences. China Daily~\footnote{China Daily is an English-language daily newspaper owned by the Central Propaganda Department of the Chinese Communist Party.} also covered a case in which an older adult in Jiangxi was nearly defrauded of two million yuan by an AI-generated virtual figure~\cite{ma2025aisafeguards}. Such influencers often adopt the personas of attractive young men or women, addressing viewers as ``sister,'' ``brother,'' or even ``mother''~\cite{xu2023jin, shen2020older}. Although they appear warm and caring, many of these accounts are ultimately commercial enterprises: by cultivating trust through prolonged interaction, they gradually encourage older adults to purchase products such as health supplements or financial services marketed as ``what the grandson uses''~\cite{legalweekly2025aitrap}.

Despite this emerging trend, systematic research on how older adults perceive and interact with AI-generated influencers remains scarce. 
Most existing literature continues to treat virtual influencers as youth-oriented~\cite{deng2023effects, li2024social}, leaving a significant gap in understanding their influence on older audiences.

%% file: section/4Methods.tex
\section{Methods}
We adopted a mixed-method approach to investigate how older adults experience social support from AI-generated influencers.
To address RQ1, we conducted an observational analysis of short-form videos featuring AI-generated influencers on Douyin and Kuaishou, focusing on the design and communicative strategies these characters employed to create a sense of social support for older viewers.
From this analysis, we identified recurring design strategies and user engagement patterns, particularly the tendency for viewers to interact with AI influencers through kinship-based role-play. These observations informed the development of our interview protocol.
To address RQ2 and RQ3, we carried out in-depth interviews with 16 older adults who had watched or interacted with AI-generated influencer content. The interviews explored their lived experiences of social support, including the types of support they perceived (e.g., emotional comfort, companionship, informational cues), their everyday practices of engagement, and their reflections on how such support compared with that from family members, friends, and human influencers.
Overall, the observation analysis provided a descriptive account of support cues and confirmed the presence of kinship-based role-play, while the interviews offered participants’ interpretations of these same patterns. Together, these methods allowed us to triangulate how design practices (RQ1) relate to older adults’ lived experiences with kinship-role AI influencers (RQ2–RQ3).

\subsection{Video Analysis}
To address RQ1, we observed short-form videos on Douyin and Kuaishou, focusing on the communicative and design strategies that conveyed social support to older adults.
Observation study is a non-intrusive research method in which researchers systematically examine naturally occurring behaviors, interactions, or artifacts without intervention or direct engagement with individuals. In social media research, this approach is widely used to analyze publicly available content~\cite{kamila2020secondary} and has been adopted in prior HCI studies~\cite{tang2022dare, wang2024there}. Because our study examines AI-generated short videos and user comment interactions as naturally occurring artifacts on short-video platforms, this method is well aligned with our research goals.
Note that, our goal in observing the videos was not to conduct statistical inference but to describe the general design and communicative patterns used by AI-generated influencers.

\subsubsection{Video Platform.}
We collected videos featuring AI-generated influencers targeting older adults from Douyin and Kuaishou, China's two largest short-video platforms. Both platforms have significant older adult user populations, with users aged 51 and above comprising 21.3\% of Douyin's 1 billion monthly active users and 21.5\% of Kuaishou's 573 million users as of March 2025~\cite{questmobile2025, questmobile2024newmedia}. However, the platforms differ in their geographic user distribution: Douyin attracts more users from Tier-1 cities, while Kuaishou has a higher percentage of users from Tier-4 cities~\footnote{China's city tier system categorizes cities by economic development and population, with Tier-1 being major metropolitan areas and Tier-4 being smaller cities and towns.~\cite{wikipedia_chinese_city_tier}}. We selected both platforms to ensure demographic diversity and comprehensive representation of Chinese short-video users.

\subsubsection{Video Collection.}

We employed an open-source social media crawler, MediaCrawler~\cite{mediacrawler2024}, to collect video data from Douyin and Kuaishou, following established practices in prior research~\cite{armin2024support, wang2024critical}.

To identify comprehensive and relevant video content, we developed a list of keywords through an iterative discovery approach. 
Beginning with the seed term "AI-generated characters", three researchers independently open-coded 500 randomly selected videos from both platforms to identify frequently associated Chinese terms. 
Through collaborative discussions, the team refined these terms and reached consensus. 
The final list includes keywords such as \textit{``AI cute child,''} \textit{``Handsome idol,''} and \textit{``Dear sister''} (full keyword list provided in the Appendix~\ref{app:keywords}).
Using these keywords, we collected the top 50 videos per keyword based on the comprehensive ranking order.

\subsubsection{Video Selection.}
Among the videos collected using our keyword search, many contained content unrelated to AI-generated influencers, such as videos featuring family members using AI filters. To ensure our dataset focused specifically on AI-generated virtual influencers targeting older adults, three researchers manually reviewed the collected videos to identify the videos featuring AI-generated influencers targeting older adults.

Selection criteria focused on the influencer's role and language-for instance, child characters addressing elderly viewers as \textit{"grandma"} or adult characters using familial terms like \textit{"sister"} or \textit{"mother."} 
Video collection and screening were conducted iteratively. 
\myhl{
After each review round, the research team discussed newly identified design features and communicative strategies. Data collection continued until additional videos no longer yielded substantively new influencer roles, communicative patterns, or interactional features relevant to kinship-role performances, indicating \textit{data saturation}.
}

From the total video pool collected via keyword searches, 224 videos met our criteria (106 from Douyin, 118 from Kuaishou) and were published between March 2022 and July 2025.
These videos included 18,650 comments (8,586 from Douyin, 10,064 from Kuaishou). 
On average, Douyin videos received 75,491 likes per video, while Kuaishou videos averaged 5,618 likes.

\subsubsection{Video Observation and Data Analysis}.
To understand the characteristics and strategies of AI-generated influencers targeting older adults, we conducted video observation~\cite{wang2024there}. 
We conducted a qualitative content analysis of both visual cues (appearance, setting, facial expressions) and conversational cues (address terms, emotional tone, narrative framing, interaction prompts). The first author open-coded an initial subset of 30 videos to identify recurring elements and discussed emerging patterns with the second author. Through these discussions, we collaboratively developed a concise codebook capturing the main kinship roles, visual design features, and conversational strategies.
We applied the codebook to the full dataset, meeting regularly to resolve discrepancies and refine category boundaries.
Visual elements were interpreted based on character age presentation, aesthetic style, professional identity cues, and background settings. 
Conversational elements were interpreted through address forms (e.g., “grandma,” “sister”), tone, emotional expressions, and platform-specific engagement prompts. 
Example analytic decisions included grouping lines such as ``light up the lamps for me'' and ``give me a little red flower'' under interaction prompts due to their shared persuasive function.
User comments were reviewed to contextualize the videos, but we did not perform systematic coding due to ethical considerations regarding user privacy. Instead, we documented only representative, de-identified examples to illustrate the interaction patterns observed.
This descriptive coding process enabled us to synthesize the three pattern groups reported in Table \ref{tab:taxonomy}, and also directly informed the design of our interview protocol.

\subsection{Interview}
To address RQ2 and RQ3, we conducted semi-structured interviews with 16 older adults, examining their experiences of support, everyday practices of engagement, and attitudes towards AI-generated influencers relative to human relationships.

\subsubsection{Participants}
We interviewed 16 participants (P1–P16; 12 female, 4 male), aged between 52 and 75 years (M = 60.44, SD = 6.43), comprising retired, semi-retired and currently employed individuals. 
Before the interviews, we confirmed that all participants had experience watching virtual influencers on short-video platforms by asking them to share their watch history from short-video platforms or provide screenshots of videos they typically viewed.
Table~\ref{tab:participants} summarises their demographic, educational, the short-video platforms they typically watch, and their self-reported frequency of watching AI-generated influencers.

Initially, we attempted to recruit participants by directly messaging them under virtual influencer videos. After contacting approximately 50 individuals over two weeks without receiving any responses, we shifted our approach. We distributed recruitment posters on social media, targeting people whose parents had experience watching virtual influencers. 
This method helped us to identify the first few participants, after which we used snowball sampling, asking these participants to recommend friends with similar experience.
In the end, we recruited participants primarily through referrals from their children or recommendations from other participants.

\newcolumntype{L}[1]{>{\raggedright\arraybackslash}p{#1}}
\newcolumntype{C}[1]{>{\centering\arraybackslash}p{#1}}

\begin{table*}[ht]
\centering
\small  
\caption{Demographic summary of interview participants. "---" indicates that the participant did not specify their information. Platform Use: D = Douyin; K = Kuaishou.}
\begin{tabular}{
C{0.3cm}  
C{0.7cm}  
C{0.4cm}  
L{1.6cm}  
L{1.2cm}  
L{1.2cm}  
L{2.2cm}  
L{2.3cm}  
L{4.5cm}  
}
\toprule
\textbf{ID} & \textbf{Gender} & \textbf{Age} & \textbf{Education} &
\textbf{Residence} & \textbf{Status} &
\textbf{Platform Use} & \textbf{Family} & \textbf{Digital Literacy} \\
\midrule
P1  & F & 61 & Vocational Secondary & Urban & Retired 
    & D, K (Often)
    & Living with spouse 
    & Never used AI before interview; Previously used computers for work tasks \\
\rowcolor{gray!10}
P2  & F & 57 & Junior High & Rural & Employed 
    & D (Daily)
    & Living with spouse and children
    & Never used AI before interview \\
P3  & M & 61 & --- & Urban & Retired  
    & D (Occasionally)
    & Living with spouse
    & Uses AI frequently (smart home assistants) \\
\rowcolor{gray!10}
P4  & F & 58 & Junior High & Urban & Retired 
    & D (Occasionally)
    & Living with spouse
    & Uses AI frequently (conversational AI) \\
P5  & F & 54 & Junior High & Urban & Retired
    & K (Often)
    & Living with grandchildren
    & Heard of AI only (robots) \\
\rowcolor{gray!10}
P6  & F & 69 & College Diploma & Urban & Retired
    & D (Often)
    & Living with grandchildren
    & Heard of AI only (digital humans) \\
P7  & M & 54 & Junior High & Rural & Employed
    & D, K (Occasionally)
    & Living with spouse and children
    & Heard of AI only (short videos explaining AI) \\
\rowcolor{gray!10}
P8  & M & 55 & Senior High & Urban & Employed
    & D, K (Occasionally)
    & Living with spouse and children
    & Uses AI occasionally (AI video generation) \\
P9  & F & 57 & Primary School & Rural & Retired
    & D (Occasionally)
    & Living alone
    & Heard of AI only (robots) \\
\rowcolor{gray!10}
P10 & F & 52 & Senior High & Urban & Retired
    & D (Often)
    & Living with children
    & Heard of AI only (Deepseek, Doubao) \\
P11 & F & 65 & Junior High & Urban & Retired
    & D (Often)
    & Living with children
    & Uses AI frequently (smart home devices) \\
\rowcolor{gray!10}
P12 & M & 56 & Junior High & Rural & Employed
    & D (Occasionally)
    & Living alone
    & Used AI occasionally (AI photo filters) \\
P13 & F & 61 & Senior High & Urban & Retired
    & D, K (Daily)
    & Living with children
    & Uses AI daily (Doubao) \\
\rowcolor{gray!10}
P14 & F & 75 & Primary School & Urban & Retired
    & D (Occasionally)
    & Living with grandchildren
    & Never used AI; Not familiar with concept of AI \\
P15 & F & 63 & Senior High & Rural & Semi-retired
    & D, K (Often)
    & Living with spouse and children
    & Uses AI frequently (Doubao) \\
\rowcolor{gray!10}
P16 & F & 69 & No formal education & Rural & Retired
    & D (Occasionally)
    & Living with spouse and children
    & Uses AI frequently (Doubao, Deepseek) \\
\bottomrule
\end{tabular}
\label{tab:participants}
\end{table*}

\subsubsection{Interview Procedure}
The interviews were conducted between June and July 2025, using online voice calls via WeChat. 
Each session lasted about one hour, and participants received 80 CNY as compensation for their time.
Interviews began with a brief introduction to the study’s purpose, followed by an informed consent process. We assured participants of their anonymity and their right to decline any questions or withdraw from the study at any time. After obtaining consent, we audio-recorded the interviews. We structured the interviews around five main parts: (1) participants’ demographic background and short-video platforms usage habits; (2) their motivations for watching AI-generated influencer videos such as "handsome young men" or "cute children"; (3) their liking, commenting behavior and interaction practices; (4) their attitudes toward AI-generated virtual influencers; and (5) any concerns or broader social reflections they had regarding these technologies. We avoided asking hypothetical questions and encouraged participants to share real-life examples and concrete experiences. 
The interviews were conducted in Chinese and recorded by the first and second authors.

\subsubsection{Interview Transcript Analysis}

We transcribed all interviews verbatim and conducted an qualitative thematic analysis to identify patterns across participants’ responses. 
We selected thematic analysis because the study aims to explore how older adults interpret kinship-role AI influencers, a phenomenon that is culturally situated and not well theorized in existing HCI literature. An inductive approach~\cite{naeem2023step} allowed themes to emerge from participants' accounts rather than being imposed by prior frameworks. While coding was primarily semantic, we also engaged in latent analysis~\cite{braun2012thematic} to interpret how participants’ descriptions were shaped by broader Chinese familial norms and expectations.
Following Braun and Clarke’s six-phase approach to thematic analysis~\cite{braun2012thematic}, we began by thoroughly familiarizing ourselves with the data through reviewing the transcripts. 
We then systematically generated initial codes by labeling meaningful segments of text that reflected participants’ experiences, perceptions, and emotional responses. 
The first and second authors independently coded the first six transcripts to surface preliminary codes and diverse interpretations. Discrepancies were discussed collaboratively to refine the codebook; in line with reflexive thematic analysis, these differences were treated as analytically generative rather than as errors requiring statistical reconciliation. Because reflexive thematic analysis does not assume mechanical consistency across coders, we did not calculate intercoder agreement~\cite{braun2021can}.
After reaching a shared interpretive framework, the first author applied the refined codebook to the remaining transcripts, with weekly team discussions throughout July 2025 to review emerging patterns and refine theme boundaries.

In the later analytic stages, we grouped and connected codes into higher-order themes by examining conditions, contexts, and consequences reflected in the data. This organizational step was informed by the relational logic of axial coding~\cite{kendall1999axial}, and themes were iteratively reviewed and refined to ensure they captured both shared patterns and the cultural nuance present in participants’ accounts.

\subsection{Ethical Considerations}
This study received approval from our institutional review board (IRB) and adhered to ethical guidelines as well as platform terms of service~\cite{townsend2017ethics}, consistent with practices in prior HCI research~\cite{armin2024support, schaadhardt2023laughing}.

For the first part of the study, our data collection focused exclusively on publicly accessible content and avoided any restricted or private areas. We extracted only openly available metadata, e.g., URLs, video IDs, publication dates, like counts, and comment counts, and deliberately excluded all personally identifiable information (e.g., usernames, avatars). Although some publicly visible content, such as video comments, was temporarily reviewed to understand typical user behaviors and inform the design of our interview guide, this material was not included in the formal analysis. Any examples shown in the paper were de-identified to prevent re-identification, consistent with established norms for responsible social media research in HCI~\cite{wang2024there, armin2024support}.

For the second part of the study, we recruited older adult participants through their adult children or friends, who confirmed that participants had prior exposure to AI-generated characters and were interested in discussing the topic. Prior to each interview, we obtained informed consent and emphasized that participation was voluntary; participants could decline any question or withdraw at any time. The analysis was based solely on interview transcripts and did not involve participants’ viewing histories or account information. All personally identifiable details were removed prior to analysis and reporting to ensure participant privacy.

\subsection{Researcher Reflexivity}
As researchers trained in HCI and social computing, our interpretations were shaped by our prior work on older adults’ digital practices, human-AI interaction, and Chinese familial norms. These backgrounds sensitized us to aspects of social support, kinship expressions, and intergenerational expectations embedded in the videos and interviews. At the same time, all members of the research team were younger and more digitally fluent than our participants, which created a generational and technological distance that could influence how we understood their emotional engagements with AI-generated influencers.

We acknowledge that these positionalities inevitably shaped how we attended to kinship metaphors, emotional expressions, and intergenerational themes in the data. While analytic safeguards such as independent coding and team discussions helped surface divergent interpretations, they do not eliminate the influence of our backgrounds. Rather, these practices served as moments for us to reflect on how our assumptions might enter the analysis and to remain attentive to grounding interpretations in participants’ accounts.

%% file: section/5Results.tex
\section{Results}
This section reports findings from both the observation study and the interviews. 
We first present the design strategies identified in the observation study (Section~\ref{sec:design-strategy}). 
We then turn to the interviews, in which older adults described their perceptions and experiences of interactions with kinship-role AI-generated influencers.
Their accounts are presented in four themes: how kinship cues shaped their sense of support (Section \ref{sec:perceived-support}), their concerns and potential risks about these interactions (Section \ref{sec:perceived-risk}), and how virtual kinship intersected with their real family relationships (Section \ref{sec:virtual-kinship-real-family}). Finally, we present an additional theme on the broader sociocultural factors that influenced their engagement with AI (Section \ref{sec:socialcultural-forces}). A summary of all themes is provided in Table \ref{tab:thematic-analysis}.

\begin{figure*}
    \centering
    \includegraphics[width=\linewidth]{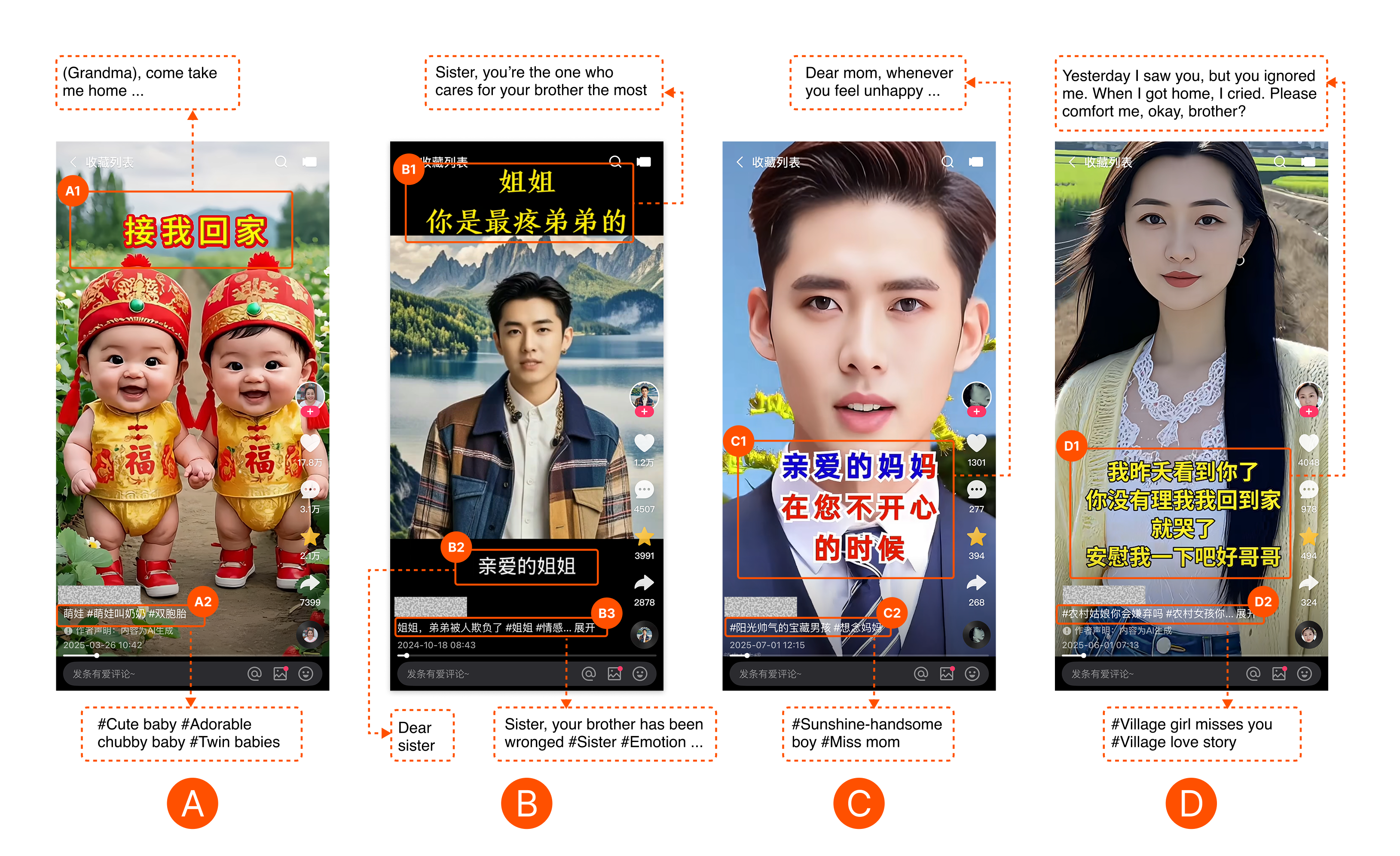}
    \caption{
    Example screenshots of AI-generated influencer videos.  
    A: \textit{AI Grandchildren} (170k+ likes, 30k+ comments, with an AI-generated content label);
    B: \textit{AI Brother} (12k+ likes, 4k+ comments, no AI-generated content label);  
    C: \textit{AI Son} (1k+ likes, 200+ comments, no AI-generated content label);  
    D: \textit{AI Sister} (4k+ likes, 900+ comments, with an AI-generated content label).
    }
    \label{fig:character-overview}
\end{figure*}

\begin{table*}[ht]
\centering
\caption{Taxonomy of design strategies used by AI-generated influencers to create social support for older adults (RQ1).}
\begin{tabular}{>{\arraybackslash}p{2cm} >{\arraybackslash}p{3cm} >{\arraybackslash}p{9.5cm}}
\toprule
\textbf{Dimension} & \textbf{Sub-categories} & \textbf{Examples} \\
\midrule
\multirow{4}{=}{Kinship Role} 
 & \cellcolor{gray!10} Grandchildren  & 
 \cellcolor{gray!10} ``We wish grandma happiness as vast as the Eastern Sea.'' \\
 & Children  & 
 ``Dear mom, whenever you feel unhappy, remember that you still have a son who cares about you.''
 \\
 & \cellcolor{gray!10} Siblings  & 
 \cellcolor{gray!10} 
  ``Dear sister, I will always support you. I have something significant to share with you...''
 \\
 & Cousins & ``Auntie, you haven’t come to see me for so long!'' \\
\hline
\multirow{5}{=}{Design}
 & \cellcolor{gray!10}  Appearance & \cellcolor{gray!10}  Human-like, cheerful, attractive  \\
 & Voice & 
 Slow, clear, natural-sounding \\
  & \cellcolor{gray!10} Professional role & \cellcolor{gray!10} Doctor, pilot, soldier, entrepreneur, actor, diplomat \\
  & Background Setting & 
  Farms, offices, airplanes \\
 & \cellcolor{gray!10} Music & \cellcolor{gray!10} Folk songs, sentimental ballads \\
\hline
\multirow{3}{=}{Conversational Strategy} 
 & Emotional expression & 
 ``Mom, I’ll always love you.'' \newline
 ``You are more important than money.'' \newline
 ``Please forgive me, okay?''
 \\
 & \cellcolor{gray!10} Storytelling &  \cellcolor{gray!10} ``If one day I came to your home, would you be willing to let me share a meal with you?'' \newline
 ``Everyone says you are a modest talent, your brilliance only grows with age \ldots''
 \\
 & Interaction prompts & 
 ``Click the little heart on the right.” \newline
 ``Light up the lamps for me.'' \newline
 ``Don’t forget to follow me.''
 \\
\bottomrule
\end{tabular}
\label{tab:taxonomy}
\end{table*}

\subsection{Design of AI-generated Influencers (RQ1)}
\label{sec:design-strategy}
In this section, we present the design of popular AI-generated characters on short-video platforms that target older adults. This analysis provides an essential context for understanding the interactions of older adults with these influencers in the following sections. We organized our findings into three themes: character role, visual and audio elements, and conversational strategy (see Table \ref{tab:taxonomy}).

\subsubsection{Kinship Role.}
Of the 224 videos we analyzed, 56 AI-generated influencers portrayed children, 165 portrayed young adults, and 3 portrayed middle-aged or older adults.
From language use, we identified family relationships when characters introduced themselves as a \textit{``son,''} \textit{``daughter,''} \textit{``younger brother,''} \textit{``younger sister,''} or \textit{``grandchild.''}

\subsubsection{Design.}
Of the 224 videos we analyzed, all AI-generated influencers featured highly human-like appearances, with no cartoon-style characters.
Childlike characters were typically portrayed as round-faced, fair-skinned, and cheerful, evoking warmth and affection (see Figure~\ref{fig:character-overview}A). 
Young and middle-aged characters were generally portrayed as attractive, with smooth skin, bright eyes, and delicate facial features (see Figure~\ref{fig:character-overview}B-D). Their expressions were typically limited to bright, cheerful smiles.
In nearly all videos, the characters spoke directly to the audience in slow, clear speech, and most synthesized voices sounded natural and easy to understand.

From these visual cues, we inferred a variety of professional identities, such as doctors, pilots, soldiers, entrepreneurs, actors, and diplomats. These roles align with occupations that are traditionally regarded as respectable, aspirational, or ``ideal'' within many Chinese family contexts.
In terms of gender, male characters exhibited a wider range of roles, spanning from young students and soldiers to mature entrepreneurs. In contrast, female characters were represented in a more limited scope, predominantly portrayed as ``girls from the countryside'' or ``elite class women.''

These roles were reinforced through clothing and setting. Their clothing ranged from colorful outfits to deliberately plain attire, evoking a simple yet familiar aesthetic. 
Characters were typically placed in settings that matched their roles: For example, ``girls from the countryside'' characters appeared in village or farmland scenes, while business-oriented characters were placed in conference rooms or airplane cabins.

Lastly, some videos included background music, often featuring folk-inspired ballads and sentimental popular songs, e.g., Wan'ai Qian'en~\footnote{\textit{Wan'ai Qian'en} is a popular ballad that reflects on childhood memories and expresses gratitude toward parents. For example, its lyrics recall ``my little hands and little feet... the laughter I left on your back... and the realization, upon returning home, that your waist has quietly bent with age.''}.

\begin{figure*}
    \centering
    \includegraphics[width=\linewidth]{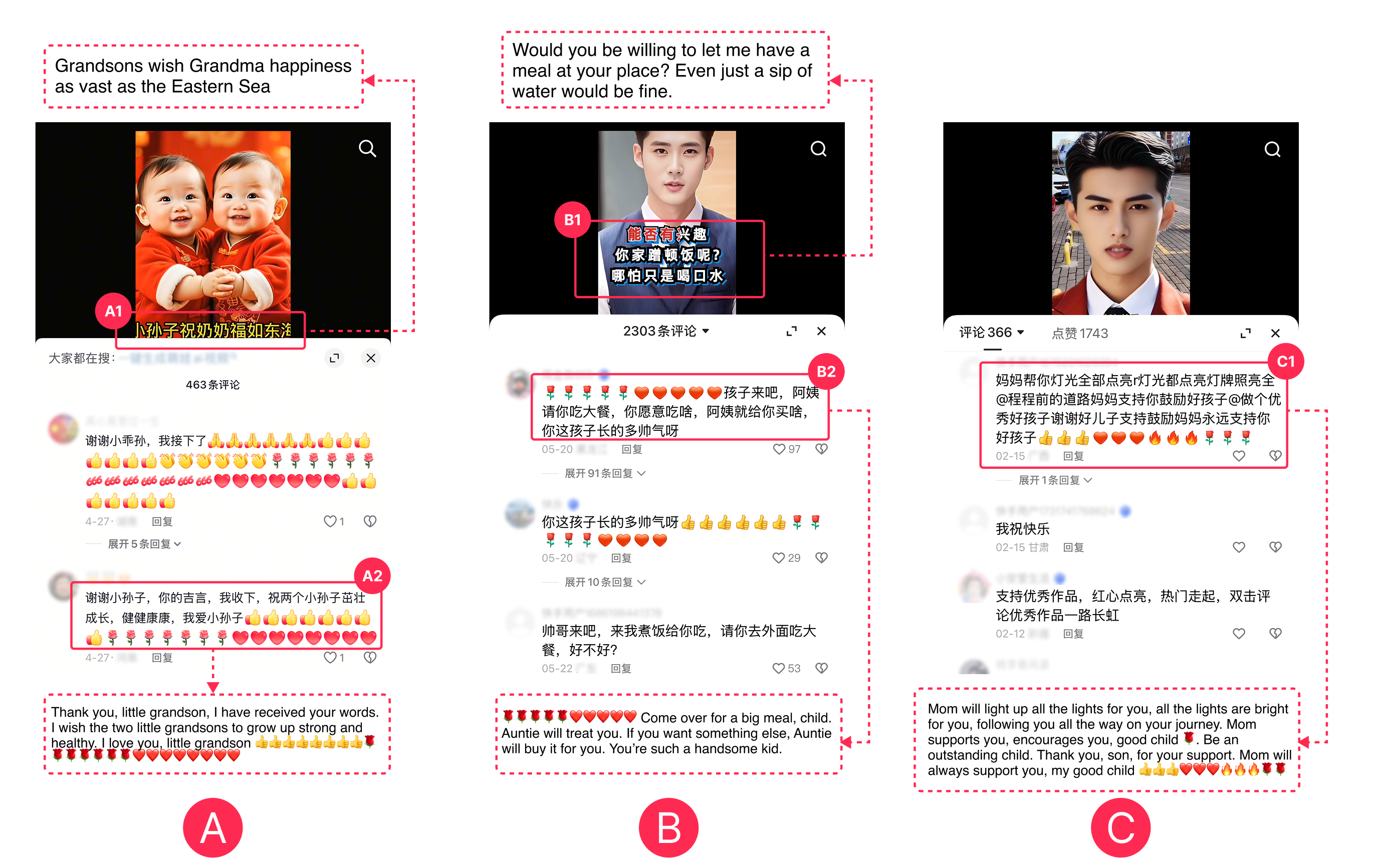}
    \caption{Example screenshots illustrating different conversational strategies in AI-generated influencer videos and how users engaged with them. A: Emotional expression, where users conveyed feelings of closeness. B: Storytelling designed to elicit sympathy, where users responded by sharing their own sympathy. C: Interaction engagement prompts, where users referenced their ongoing support in platform-specific ways.}
    \label{fig:engagement-example}
\end{figure*}

\subsubsection{Conversational Strategy.}
\newcounter{step}
\setcounter{step}{1}
Our analysis revealed three recurring conversational strategies across the AI-generated influencer videos (see Appendix~\ref{app:example-scripts} for more illustrative examples):

\Roman{step}) First, many influencers employed emotional expression in their communication.
For example, an \textit{AI son}’s script read: \textit{``Dear Mom, whenever you feel unhappy, remember that you still have a son who cares about you. Today I am in a perfect mood and want to surprise you. Mom, your son will always love you.''} (see Figure~\ref{fig:character-overview}C).
Beyond these direct declarations of love, the AI-generated influencers consistently conveyed blessings and affectionate well-wishes, such as \textit{``I wish you health and happiness.''} (see Figure~\ref{fig:engagement-example}A). For example, an \textit{AI younger brother} addressed his audience by saying:
\begin{quote}
\textit{
``Dear sister, I will always support you. I have something significant to share with you [...] I want to send you my best wishes. Recently, I have had a very strong feeling that you will encounter something completely unexpected in the coming days. But no matter what happens, I hope you can face it bravely [...]''
}
\end{quote}

\stepcounter{step}
\Roman{step}) The second strategy involved using fictional stories or scenarios designed to elicit sympathy. For example, some characters expressed loneliness (e.g., \textit{``I feel so lonely, could you comfort me?''}) or recounted imagined experiences to provoke care and concern from the audience (e.g., \textit{``If one day I came to your home, would you be willing to let me share a meal with you? Even just a sip of water would be enough.''}) (see Figure~\ref{fig:engagement-example}B). One AI younger brother’s script, for instance, narrated: 
\begin{quote}
    \textit{``Someone called me and told me not to bother you anymore because you already have him. I was stunned, standing in the pouring rain, overwhelmed with sadness. I had so much to say but didn’t know how to begin. I wanted to leave this place filled with memories of you—perhaps then my heart would not hurt so much. I only hope that both of us can live well in the future.''}
\end{quote}

\stepcounter{step}
\Roman{step}) 
Finally, the influencers frequently employed interaction prompts, encouraging viewers to like, follow, or favorite their videos. These actions were often framed in romantic or poetic language, such as describing a ``like'' as giving a flower or a ``favorite'' as lighting a small lamp. For instance, an \textit{AI grandson} prompted viewers to comment by saying: \textit{``Hello Grandma, today is your grandson’s birthday. Could you give me a little red flower to encourage me? I know you love me the most. Love you!''} Similarly, referring to the star-shaped favorite icon, which resembles a small light, one AI-generated influencer remarked: \textit{``Can you light up all the lamps for me? Then we will not get lost.''} (see Figure~\ref{fig:engagement-example}C).

\begin{table*}[ht]
\centering
\caption{Summary of Thematic Analysis}
\renewcommand{\arraystretch}{0.9}
\begin{tabular}{>{\arraybackslash}p{5cm} >{\arraybackslash}p{10cm}}
\toprule
\textbf{Theme} & \textbf{Subtheme} \\
\midrule
\multirow{4}{=}[-1ex]{Social-Psychological Needs Fulfilled Through Virtual Kinship}
 & Fulfilling Emotional Needs Through Kinship-based Addressing \\
 & \cellcolor{gray!10} Fulfilling Cognitive Needs Through Shared Generational Interests  \\
 & Fulfilling Moral and Cultural Alignment Needs Through Values \\
 & \cellcolor{gray!10} Fulfilling Relational Needs for Presence and Companionship Without Burden  \\
\midrule
\multirow{3}{=}[-1ex]{Concerns and Risks from Virtual Kinship}
 & AI Impersonation Scams Using Family Member-Like Appearances  \\
 & \cellcolor{gray!10}  Emotional Displacement Toward Kinship-Role AI Influencers \\
  & Overtrust Toward Kinship-Role AI Influencers Viewed as Non-Commercial \\
\midrule
\multirow{2}{=}{Virtual Kinship in Tension with Real Family Ties}
 & \cellcolor{gray!10} Bridging Intergenerational Gaps Through Perspective-Taking  \\
 & Social Comparison Between AI Kinship Personas and Real Family Members \\
\midrule
\multirow{2}{=}{Sociocultural Forces Shaping Virtual Kinship}
 & \cellcolor{gray!10} Keeping Up With Technological Change and Expressing Collective Pride in AI Adoption \\
 & Navigating Online Ageism Through Virtual Human Avatars \\
\bottomrule
\end{tabular}
\label{tab:thematic-analysis}
\end{table*}

\subsection{Social-Psychological Needs Fulfilled Through Virtual Kinship}
\label{sec:perceived-support}

In this section, we present how older adults experienced support and benefits from kinship-role AI-generated influencers. 
Consistent with prior research~\cite{lou2023authentically}, many interviewees described the videos as a source of entertainment and relaxation (P1–3, P6, P7, P10, P11). 
The youthful vitality embodied by AI influencers often helped lift their mood in daily life (P8, P10).
However, interviewees also described experiences that extended beyond general parasocial interaction. In line with RQ2, we interpret these experiences as forms of social support, consistent with HCI and gerontology definitions that view emotional, cognitive, moral, and relational fulfillment as key dimensions of perceived social support.
We detailed the sub-themes below:

\subsubsection{Fulfilling Emotional Needs Through Kinship-based Addressing.}
Many interviewees described kinship-based addresses as a key source of emotional fulfillment. Such interactions often made them feel happy, comforted, and emotionally ``lifted.'' This resonance was especially strong when AI influencers used intergenerational cues that mimicked the tone and affection of real grandchildren. 
As P11 expressed, \textit{``Hearing those words come from a child - it’s especially endearing. It sounded so innocent that I couldn’t help but smile and say thank you.''}
Some interviewees even responded in corresponding kinship roles. For instance, P2 affectionately referred to two AI ``grandchildren'' as \textit{``my little baby.''} She recalled a moment when the AI children told her, \textit{``Grandma, you are so blessed,''} which immediately brightened her mood and prompted her to respond: \textit{``My little baby is so adorable.''}

\subsubsection{Fulfilling Cognitive Needs Through Shared Generational Interests.}
Beyond emotional resonance, many interviewees described how kinship-role AI influencers engaged with topics that aligned with their generational knowledge and everyday cognitive concerns, such as health topics, traditional remedies, historical memories, or ways of handling family dynamics. 
These themes were deeply relevant to their stage of life and were often topics their adult children did not discuss or fully understand.
P3 appreciated that AI influencers \textit{``shared lifestyle tips and recipes suitable for older people,''} which felt directly meaningful to his daily routines. 
Similarly, P15 noted that the historical content presented in the AI influencer’s videos helped her recall and articulate memories from her youth -experiences that were difficult to communicate to her children because, as she explained, \textit{``they were never in that history...they don’t know what that time looked like.''} 
Such moments made interviewees feel cognitively understood and supported in ways that real intergenerational interactions often did not provide.

\subsubsection{Fulfilling Moral and Cultural Alignment Needs Through Filial Values.}
Many interviewees emphasized that the AI personas embodied moral qualities they deeply valued, particularly filial respect and attentiveness to elders. These portrayals resonated with their expectations of how younger family members ``should'' behave, reflecting long-standing cultural scripts of filial piety. 
For instance, P11 described feeling moved when an ``AI grandson'' appeared cooking for his grandmother, explaining that the character was \textit{``especially filial and very good to his grandmother.''} Likewise, P10 appreciated scenes in which AI characters helped older adults, noting that such behavior aligned with her belief in promoting \textit{“positive energy,”} a term in the Chinese context referring to uplifting and morally affirming conduct. 
For many interviewees, such depictions represented idealized expressions of intergenerational virtue that they rarely encountered in real family interactions.

\subsubsection{Fulfilling Relational Needs for Presence and Companionship Without Burden.}
Many interviewees described kinship-role AI influencers as offering a form of companionship that felt emotionally present yet more accessible than interactions with real family members. They emphasized that AI personas were consistently responsive, warm, and patient—qualities that made the interactions feel comforting during moments of stress or solitude. As P13 noted elsewhere, \textit{``When you're unhappy, it makes you happy, it comforts you… With real people, you can’t always expect that level of response.''}
At the same time, interviewees were keenly aware that these relationships were fundamentally one-sided. This asymmetry, particularly the absence of obligation or reciprocity, made the companionship feel ``lighter'' and easier to engage with than real kin relationships.
As P10 explained, \textit{``Even if you like them (AI influencers), so what? They won’t really come to your side.''} Rather than viewing this as a limitation, interviewees often saw it as enabling. The lack of interpersonal demands allowed them to enjoy connection on their own terms. As P5 put it, \textit{``It’s fine to give emotions to AI because it belongs to you. But if you don’t want to, you don’t have to—because it’s not a real person.''}

\subsection{Concerns and Risks from Virtual Kinship}
\label{sec:perceived-risk}
In this section, we summarize the key concerns and risks associated with kinship-role AI influencers. Interviewees expressed explicit concerns about AI impersonation scams that mimic family members' appearances, as well as worries about emotional displacement that might affect real family relationships. 
We also identified an implicit risk, where many interviewees perceived these AI characters as non-commercial and emotionally dependable, which contributed to a tendency to overtrust them.
We elaborate on these concerns below:

\subsubsection{AI Impersonation Scams Using Family Member-Like Appearances.}
A recurring concern was the risk of deception from increasingly realistic AI-generated content (P3, P4, P6, P13). 
Several interviewees feared that the lifelike appearance of these virtual figures could be exploited by scammers, particularly if AI were used to imitate familiar people such as their children. As P13 explained: \textit{``Because some of our phone numbers, or WeChat or Douyin accounts could leak videos of our family members... I'm worried that someone could use them to clone one of my family members and scam me. This really concerns me.''}

\subsubsection{Emotional Displacement Toward Kinship-Role AI Influencers.} 
A few interviewees also worried that relying too heavily on AI-generated influencers might come at the expense of real family relationships. They feared that emotional investment in digital figures could gradually erode intimacy with children or spouses, creating distance within the household. P5, for instance, explained why she was hesitant to embrace AI for deeper emotional connections: \textit{``If you’re single, it’s quite good ... it can serve as a kind of emotional outlet. But for people who already have a family, I think it’s not so good. It feels like you might invest some of your emotions into it ... just like being distracted.''} Her reflection illustrates how, even as interviewees recognized the comfort AI could provide, they remained cautious about the risk of diverting emotional energy away from authentic family ties.

\subsubsection{Overtrust Toward Kinship-Role AI Influencers Viewed as Non-Commercial.}
A notable subtheme was interviewees' perception that kinship-role AI influencers were more trustworthy than human influencers. 
Many interviewees believed that, unlike human creators who might be driven by commercial interests or product promotion, AI characters had no underlying profit motive and therefore felt ``purer'' and more reliable as sources of emotional support. As P13 explained, \textit{``With digital figures, there’s no profit involved. With real influencers - whether they are vloggers or people doing product promotions - you know there's always some interest behind it. But with digital figures, it feels more genuine, more pure.''} This perceived absence of commercial intent contributed to a tendency to trust these AI personas, with several interviewees noting that they had purchased products recommended by the AI influencers.
Although interviewees did not identify this as a direct concern, such lowered vigilance represents an implicit risk within virtual kinship interactions.

\subsection{Virtual Kinship in Tension with Real Family Ties}
\label{sec:virtual-kinship-real-family}

In this section, we focus on how the role of "kinship" of AI influencers have been used to influence the real family ties, including both positive and negative ways.

\subsubsection{Bridging Intergenerational Gaps Through Perspective-Taking.}
For some interviewees, kinship-role AI personas offered explanations and reframing that helped them view family conflicts from a different angle, easing misunderstandings and reducing emotional escalation. For example, P15 described how the AI’s sharing of handling interpersonal tensions enabled her to approach a family dispute with greater empathy and composure: \textit{``Because I talked it over with the AI, I knew how to deal with the issue better. When I later faced my family, the problem was already smaller.''} 
Similarly, P3 reflected that such interactions promoted mutual understanding within family relationships, describing the experience as \textit{``a process of building understanding between family members.''} 
Through these forms of cognitive and emotional reframing, kinship-role AI influencers helped some older adults bridge intergenerational gaps by encouraging a calmer, more understanding approach to real-life family interactions.

\subsubsection{Social Comparison Between AI Kinship Personas and Real Family Members.}
Interviewees often portrayed AI-generated influencers as flawless in appearance and behavior, sometimes leading interviewees to compare them to their real-life family members. 
Such comparisons reveal the fragile boundary between supportive affirmation and unrealistic expectations, as P10 indicated: \textit{``When I see a handsome AI character, I can’t help but imagine turning my husband into that type. They just look so stylish, and I wish he could be like that too.''}  
Others reflected more critically. For example, P3 compared AI children with his grandchildren: \textit{``These AI children are so sweet and obedient. I can’t help comparing them to my real grandkids.''} He cautioned that such portrayals could create unfair expectations: \textit{``These AI kids behave in ways that might never be possible in real life, so it’s not fair to expect real children to follow them.''}

\subsection{Sociocultural Forces Shaping Virtual Kinship}
\label{sec:socialcultural-forces}
In the final section, we describe how interviewees’ engagement with virtual kinship was shaped by broader social and cultural contexts, including collective pride in technological advancement, a perceived need to keep pace with rapid technological change, and experiences of online ageism that prompted them to seek alternative ways of participating online.

\subsubsection{Keeping Up With Technological Change and Expressing Collective Pride in AI Adoption.}
Several interviewees described their engagement with kinship-role AI influencers as a source of pride and even honor, linking these technologies to China's broader trajectory of scientific and technological advancement. They viewed such AI systems as reflective of national progress. P14, for instance, remarked that seeing human-like AI characters evoked \textit{``a spontaneous sense of pride,''} interpreting them as evidence of the country’s scientific strength. P6 similarly explained, \textit{``People our age happen to catch the end of a technological development era. I feel quite honored. But we can’t fall behind, right?''} Others echoed this sentiment, expressing emotional reactions tied to collective achievement.
Even interviewees who felt neutral or hesitant toward AI articulated similar narratives. As P12 noted, \textit{``Even if you don’t like it, there’s nothing you can do. Society has already developed to this point. You can’t resist it… It’s all inevitable.''} Phrases such as \textit{``you have to follow the tide of the times''} were commonly invoked to describe AI adoption as both an individual necessity and a social expectation.

\begin{figure}
    \centering
    \includegraphics[width=\linewidth]{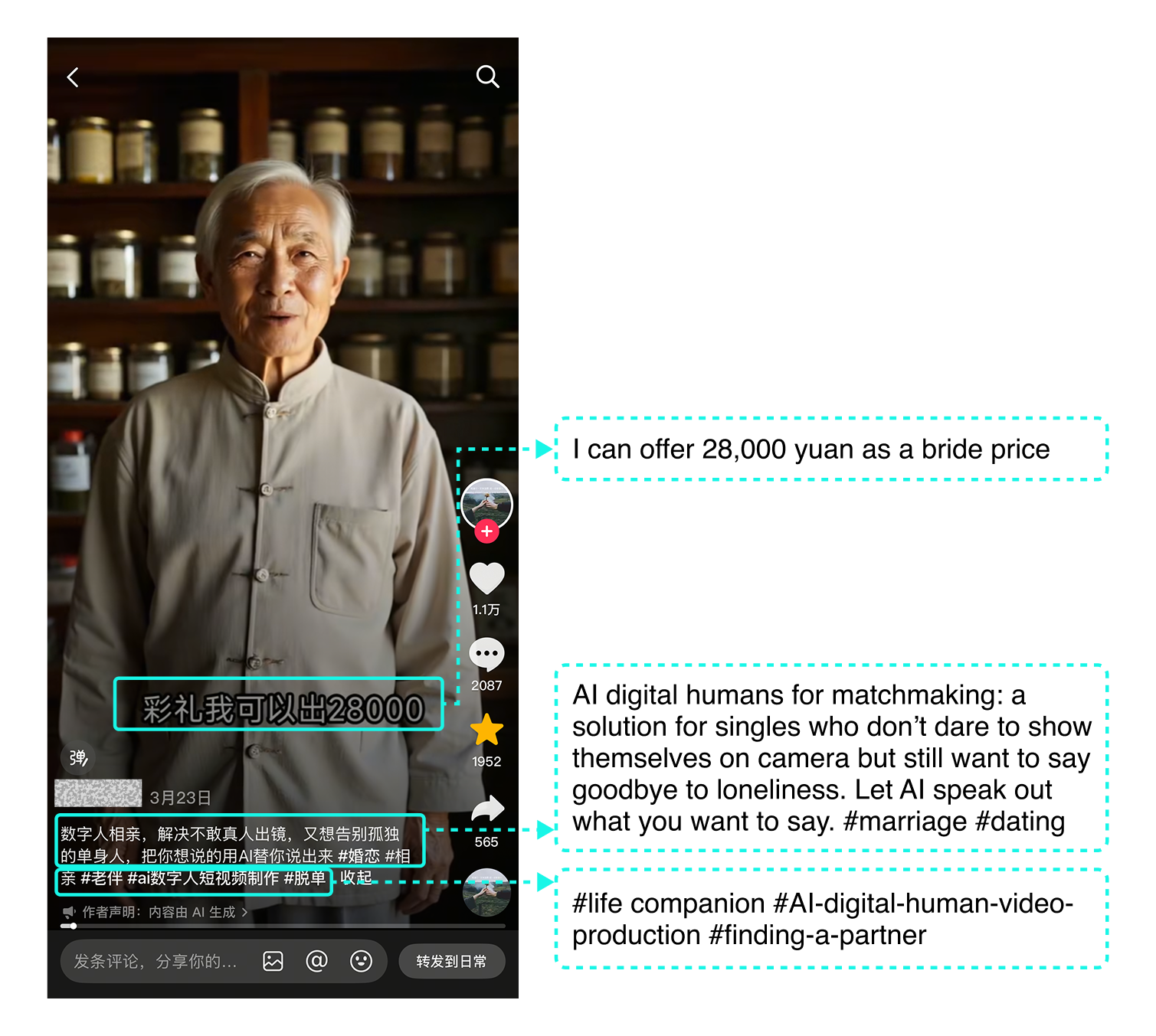}
    \caption{Example of an AI-generated influencer designed with an older adult’s appearance. In this case, the video advertises matchmaking services for older adults.}
    \label{fig:ai-self}
\end{figure}

\subsubsection{Navigating Online Ageism Through Virtual Human Avatars.}
\label{counter-digital-agesim}
Several interviewees described feeling anxious about appearing online due to ageist ridicule and negative comments, which motivated them to seek ways of participating digitally without showing their real faces. P13, for example, recounted her hesitation to appear in videos: \textit{``Some trolls criticize older people using beauty filters online, saying things like, `You’re so old, why are you even doing this?' It’s hurtful, and I no longer dare to show my face.''} Anticipating similar judgments, others expressed fear about posting or being on camera (P10–13). As P10 explained, despite her strong desire to share videos online, \textit{``I'm afraid others will laugh at me.''}
In this context, the rise of kinship-role AI influencers was perceived as especially appealing because it offered a way to participate online without exposing oneself to such judgment.
Interviewees also emphasized that this trend was far from niche: 
some AI influencers were already designed to resemble older adults, and advertised with messages such as, \textit{``If you don't want to appear on camera, you can use our AI digital human.''} (see Figure \ref{fig:ai-self}).

%% file: section/6Discussion.tex
\section{Discussion}

\subsection{Older Adults’ Online Engagement and the Role of Virtual Personas}
Older adults’ engagement in online communities is increasingly central to their social support and emotional well-being~\cite{waycott2019building, baez2019technologies} and has become a key focus within HCI research on aging~\cite{huang2025designing, knowles2019hci, knowles2024hci}. Socioemotional Selectivity Theory argues that older adults prioritize emotionally meaningful, low-conflict interactions over exploration~\cite{carstensen1992social}. This affective focus shapes how they select and interpret relational cues online, making it essential to understand how AI-generated virtual personas are perceived from their life-stage perspective.

Our findings advance understanding of how older adults interpret AI-generated virtual avatars by providing empirical accounts of their engagement with kinship-role AI influencers, a rapidly expanding category of AI generated virtual personas that is gaining visibility and influence on Chinese short video platforms such as Douyin and Kuaishou~\cite{legalweekly2025aitrap, thepaper2025aigrandson}.
\myhl{
Interviewees commonly described these figures as entertaining or relaxing, consistent with prior work on parasocial interaction that highlights feelings of familiarity and emotional connection toward media figures in the absence of reciprocal relationships~\cite{arsenyan2021almost, stein2024parasocial}. At the same time, participants also described how AI personas adopting explicit kinship roles could resonate with broader social and emotional concerns in later life.
For example, terms of address like ``grandma,'' consistently positive affirmations, and carefully crafted conversational and visual content fostered emotional, moral, and cognitive resonance, aligning with older adults’ preferences for warmth, understanding, and piety. 
}
These dynamics help explain why such AI influencers adopting kinship roles are particularly appealing for aging populations whose social networks often narrow and who increasingly seek emotionally meaningful interactions.

A second contribution of our study concerns the role of online ageism in shaping older adults’ engagement with virtual personas~\cite{diaz2018addressing, zou2024mitigating}.
In our study, several interviewees reported experiencing ridicule or judgment when posting their own content on short-video platforms, leading to anxiety about public participation. 
For these individuals, AI avatars with human-like appearances were seen as a promising way to engage online without fear of stigma. 
Future work could build on this direction by designing personalized AI avatars that support more inclusive digital participation for older adults.
Together, these insights connect our work to ageing HCI~\cite{knowles2019hci, knowles2024hci, righi2017we} by showing how older adults' social-psychological needs intersect with broader cultural and technological contexts to shape their engagement with, perception of, and interpretation of AI-generated figures.

\subsection{Relational Mechanisms Underlying the Emergence of Virtual Kinship}
A central contribution of our study is to explain why and how AI-generated influencers adopting kinship roles prompt older adults to interact with them as if they were family members. 
Prior work on \textit{fictive kinship} shows that kinship ties can be formed through social practices and emotional connection rather than biological relatedness alone~\cite{pierre1975all, carsten2000cultures}. Building on this perspective, we argue that the kin-like relationships observed in our study extend this logic into a digitally mediated context. 
We describe this phenomenon as \textit{virtual kinship}: a relational interpretation in which AI-mediated interactions are understood through the affective and cultural norms of kinship.

This phenomenon also aligns with theoretical accounts of how and why family relationships carry distinctive emotional weight in later life. Attachment theory highlights that familial bonds are experienced as more normative, stable, and less substitutable than friendships or instrumental relationships~\cite{bowlby1982attachment}. Likewise, SST suggests that as people age, they increasingly prioritize interactions that are emotionally meaningful, predictable, and supportive over exploratory or informational goals~\cite{carstensen1992social}. AI personas that adopt kinship roles align closely with these motives, offering consistent emotional affirmation with minimal interpersonal risk.

This helps explain how virtual kinship is produced through the relational cues we identified in Section \ref{sec:perceived-support}. First, kinship-based forms of address (e.g., “grandma”) provided immediate emotional affirmation and a sense of familial closeness. Second, shared generational interests in stories, memories, or everyday concerns created cognitive resonance, making the AI feel relatable and attuned to their life world. Third, consistent alignment with filial, moral, and culturally valued behaviors~\cite{xu2007chinese} strengthened the sense of moral familiarity and trust. 
From an SST perspective, these forms of predictable, affectively positive engagement satisfy older adults’ socioemotional goals~\cite{carstensen1992social}; from an attachment perspective, the AI's stability and non-confrontational nature position it as a quasi–secure base, reinforcing kin-like interpretations~\cite{bowlby1982attachment}.

The discussion of virtual kinship offers both conceptual and design insights. Conceptually, our findings extend human-AI relationship research, which has largely focused on friend- or assistant-like roles~\cite{antony2023co, liu2025toward}. Sociology and gerontology show that family ties carry stronger norms of obligation, emotional commitment, and non-substitutability than other relationships~\cite{bowlby1982attachment, bengtson1991intergenerational}. By demonstrating how older adults interpret AI personas through these kinship logics, we broaden the relational frameworks used to understand human-AI interaction.
For design, the relational mechanisms underlying virtual kinship suggest opportunities for AI systems that support older adults~\cite{jin2024exploring, waycott2019building}. 
Although our findings arise from short-video parasocial interactions, these mechanisms may generalize to more interactive modalities, such as AI companions or digital support tools. Designing with attention to kinship-related expectations and emotional norms may help create systems that feel more resonant and trustworthy to older adults.

\subsection{Situating Virtual Kinship Within Contemporary Chinese Family and Cultural Norms}
Our findings from Section \ref{sec:perceived-support} show that the appeal of kinship-role AI influencers is shaped not only by advances in AI impersonation technology but also by the cultural and socioeconomic realities of contemporary China. 
A critical lens to interpret such responses is the cultual background of Chinese family relationship: filial piety, mutual support, and emotional attentiveness have long structured Chinese family life~\cite{xu2007chinese, carsten2000cultures, stafford2000chinese}. When AI personas draw on familiar cultural cues, such as expressions of filial concern or everyday attentiveness, they invite older adults to interact with them through the lens of family, reinforcing emotional resonance rooted in traditional culture. These insights suggest that future AI companionship designs for older adults may benefit from incorporating culturally grounded symbols of care~\cite{huang2025designing}, while remaining attentive to the ethical implications outlined above.

However, our findings in Section \ref{sec:socialcultural-forces} point to another critical lens for interpreting these dynamics: the broader policy environment and the rapid demographic and structural shifts taking place in China~\cite{que2024filial}. 
Many older adults in our study grew up in large, interdependent households where collectivist values were deeply embedded. Their adult children, by contrast, belong largely to the one-child-policy generation (1979–2015) and now live in much smaller family units~\cite{zhao2023china}. These shifts have two important consequences. First, younger adults tend to adopt more individualistic orientations, which can create gaps in how the two generations understand family roles and obligations. Second, the reduced family size concentrates caregiving responsibilities on a single child, intensifying the pressures they face in supporting aging parents~\cite{bai2019whom, que2024filial}. 
As a result, gaps in intergenerational expectations arise not only from cultural differences but also from structural conditions that cannot be resolved at the individual or household level.

Together, these structural and cultural transformations help explain why kinship-role AI influencers resonate so strongly with Chinese older adults: they symbolically reproduce forms of familial attentiveness that have become difficult to maintain under contemporary conditions. For future directions in HCI, this suggests that AI systems designed for aging populations must be understood not only as technical artifacts but also as mediators within evolving family ecologies~\cite{huang2025designing}. Designing supportive AI for older adults in China, and in other rapidly aging societies, requires sensitivity to both cultural values and the structural constraints shaping intergenerational care.

\subsection{Ethical Reflection: Algorithmic Intimacy, Emotional Displacement, and Platformized Care}
\myhl{
Algorithmic intimacy has been associated with a range of ethical concerns that may place users’ well-being at risk. In media contexts characterized by parasocial interaction, such as watching influencer videos, prior work has raised concerns about increased social isolation, excessive emotional investment, and media addiction~\cite{hartmann2016parasocial}.} 
With the emergence of more interactive technologies, such as conversational agents and AI companions, research on human–AI relationships has further highlighted risks, including manipulation and emotional dependency~\cite{zhang2025dark, li2024finding, kran2025darkbench}, where AI systems cultivate closeness through scripted warmth, personalization, and constant availability. Building on this line of work, our findings (Section~\ref{sec:perceived-risk}) point to an additional ethical concern discussed in studies of algorithmic intimacy~\cite{elliott2022algorithmic, brooks2021artificial, wiehn2023algorithms}: emotional displacement.

Unlike celebrity-based parasocial relationships, kinship metaphors map onto actual family roles. When AI ``children'' or ``siblings'' begin to serve as emotional stand-ins, they may subtly redirect affective investment away from real family members, reshaping expectations of care and intimacy. Such displacement may change how older adults interpret family relationships, reduce tolerance for ordinary intergenerational imperfections, or create emotional demands that real relatives cannot satisfy.

This concern also resonates with emerging debates about ``AI afterlife,'' in which people seek digital replicas of deceased family members~\cite{morris2025generative, lei2025ai}. Our observations of \textit{virtual kinship} point to a parallel yet distinct issue: even when relatives are alive but simply unavailable, some older adults may naturally turn to AI as a substitute for family. This raises difficult questions: What does it mean to replicate kinship roles that carry deep cultural obligations? And at what point do such systems begin to reconfigure, rather than merely supplement, human family relationships?

These risks become even more complex when considered within the broader context of platformized care~\cite{celebi2025platformization, korn2025informal}, where digital platforms increasingly mediate and monetize care services. In our study, kinship-role AI influencers are currently free to watch and engage on short video platforms, and their primary function is often commercial advertising that targets vulnerable older adults.
However, as AI companionship becomes further integrated into platform economies, these virtual ``family members'' could easily shift into paid services. Companies may begin producing personalized AI companions at scale and charging users for ongoing emotional support, in ways similar to Otome game~\footnote{An otome game is a story-based romance video game targeted towards women with a female protagonist as the player character.} but designed for older adults~\cite{lei2024game, huan2022female}.

In such a scenario, expressions of care risk becoming fully commodified, which makes the boundary between emotional support and algorithmic persuasion increasingly unclear. These developments raise important ethical questions about whether familial roles should be commercialized through AI technologies, and how platforms might exploit emotional needs while presenting such interactions as forms of care.

\subsection{Limitation \& Future Work}

Our study has several limitations.
First, our strategy of collecting the top videos per keyword based on comprehensive ranking introduced a potential popularity bias. While this approach was practical for managing a large dataset, it likely overrepresented highly visible or widely shared content, while overlooking less popular videos that may nonetheless shape older adults’ experiences. As a result, our findings may reflect the dynamics of mainstream content more than the diversity of niche or marginalized voices.
Future work could address this limitation by conducting a more comprehensive large-scale social media analysis of kinship-role AI personas targeting vulnerable groups, which may reveal design strategies and communicative patterns that are less prominent in top-ranked videos.

Second, although we included participants from different regions of China, the overall sample size was modest and skewed toward individuals with relatively stable technological access. 
Because we sought older adults who had direct experience with AI-generated influencers, we relied on purposeful sampling~\cite{suri2011purposeful} rather than aiming for representativeness.
This approach, however, limits our ability to capture how older adults with less technological experience might interpret kinship-role AI characters. 
Future work should collaborate with community organizations to recruit older adults with lower literacy, limited device access, or those in low-resource environments, enabling a broader range of perspectives to be included.

\myhl{
Third, our study does not include a direct comparison between kinship-role AI-generated influencers and AI influencers adopting non-kinship or generic roles.
This is because our study is an exploratory investigation of an emerging social phenomenon, focusing on how kinship-role AI influencers are designed and experienced by older adults in real-world contexts on short-video platforms such as Doyin and Kuaishou.
As a result, our findings should not be interpreted as evidence of the relative effectiveness or preference of kinship-based roles over other AI persona designs. 
However, this focus also limits the generalizability of our findings to broader claims about AI companionship or agent role preferences among older adults.
Future research could adopt comparative or experimental designs to examine how different agent role framings shape older adults’ perceptions, emotional responses, and engagement patterns, and to identify which aspects of kinship-based designs meaningfully differentiate them from other AI interaction paradigms.
}

\myhl{
Fourth, recruiting participants through their adult children further narrowed the representativeness of our sample. While this strategy helped us reach eligible participants, it biased the pool toward older adults who receive familial support in navigating digital platforms. As a result, our findings primarily reflect the perspectives of those who are socially connected, offering more limited insight into digitally isolated or unsupported older adults. 
Although we encourage future studies to address this gap, purposefully sampling older adults with limited social or family support remains a methodological challenge, as their limited visibility in community settings and lower engagement with digital platforms make conventional recruitment channels ineffective.
Addressing this challenge will likely require sustained collaborations with community stakeholders and more inclusive outreach strategies capable of reaching digitally marginalized populations.
}

Finally, the long-term impacts of emotionally supportive AI remain unclear. While many interviewees described positive short-term emotional effects, we do not know how persistent engagement with emotionally supportive AI agents may affect users’ mental health, relational satisfaction, or social networks over time. Longitudinal studies are needed to examine both the psychological benefits and the potential risks of emotional dependence, especially as AI systems become more persuasive, personalised, and emotionally responsive.

%% file: section/7Conclusion.tex
\section{Conclusion}
This paper examined how older adults experience social support from AI-generated influencers on short-video platforms that adopt kinship-based roles. Through analysis of 224 popular videos, we identified key design strategies employed by these influencers to make them particularly appealing to older audiences. These strategies include kinship-like address terms, visual and conversational cues, and emotionally affirming expressions.
Interviews with 16 older adults, most of whom had family support in navigating digital technologies, further revealed how these strategies translated into lived experiences of what we describe as \textit{virtual kinship}. Participants reported strong emotional resonance and perceived forms of comfort, companionship, and affirmation. At the same time, they voiced concerns about emotional displacement, as well as heightened vulnerability to misinformation, impersonation, and scams.
These experiences shaped how older adults interpreted AI personas in relation to their real family members. While kinship-based AI influencers were sometimes perceived as temporarily filling intergenerational gaps when family members were unavailable, they also prompted social comparison and highlighted tensions between idealized kinship scripts and the realities of contemporary family life.
Our findings extend HCI research by documenting this emerging phenomenon of virtual characters leveraging kinship-based roles, analyzing the design and conversational strategies involved, and illustrating how older adults perceive and negotiate these interactions. We further show how attitudes toward virtual kinship are shaped by cultural expectations and social conditions in contemporary China.
Based on our findings, we argue that designing for aging populations requires balancing low-barrier access to support with safeguards that prevent emotional displacement and over-reliance. 
Future work should critically investigate how kinship-based AI companions can complement rather than substitute real-life relationships, while promoting transparency, cultural sensitivity, and inclusive participation for older adults in digital environments.

%% file: section/8Appendix.tex
\section{Appendix}
\subsection{Keywords}
\label{app:keywords}

(AI digital human) OR 
(AI cute child) OR 
(AI child) OR 
(emotional quotes) OR 
(dear sister) OR 
(sister, brother really misses you) OR 
(handsome idol) OR 
(in the vast sea of people, meeting is fate) OR 
(handsome younger brother) OR 
(Brother Dong) OR 
(Brother Jianguo)

\subsection{Example Scripts of AI-generated Influencer}
\label{app:example-scripts}
To illustrate the typical interactional styles of AI-generated influencers, we provide several representative excerpts below. Each excerpt highlights a distinctive rhetorical strategy.

\paragraph{Caring with interaction prompt.}
\begin{quote}
    \textit{Dear sister, are you looking at your phone right now? Don’t read while lying down, it’s bad for your eyes. Remember you said you liked that song? I stayed up all night to learn it for you. Sister, can you light up the row on the right for me? Thank you so much, my dear sister.}
\end{quote}

\paragraph{Pleading and coercive tone.}
\begin{quote}
    \textit{If you keep scrolling, I’ll really be angry. I’ve called you so many times and you still ignore me. If you don’t respond today, I swear I won’t bother you again. Can you please give me just one warm hug?}
\end{quote}

\paragraph{Poetic and sentimental language.}
\begin{quote}
    \textit{I have always longed to know how you are doing. My yearning for you is like the grass longing for the spring breeze. At last, I have found you on this platform, and I deeply feel it is destiny—perhaps even a bond from a previous life—that has brought us together again. The moment I first saw you, I was overwhelmed with an indescribable excitement, as if we had met before in some familiar scene \ldots}
\end{quote}